\documentclass[twocolumn,aps,prl,showpacs]{revtex4}

\usepackage{amsmath} 
\usepackage{graphicx}
\usepackage{bm}
\usepackage{amsfonts}

\begin{document}

\title{Glassy dynamics of partially pinned fluids: an alternative mode-coupling approach}

\author{Grzegorz Szamel and Elijah Flenner}

\affiliation{Department of Chemistry, 
Colorado State University, Fort Collins, CO 80523
}

\date{\today}


\begin{abstract}
We use a simple mode-coupling approach to investigate glassy dynamics of 
partially pinned fluid systems. Our approach is different from the mode-coupling theory 
developed by Krakoviack [Phys. Rev. Lett. \textbf{94}, 065703 (2005), 
Phys. Rev. E \textbf{84}, 050501(R) (2011)]. In contrast 
to Krakoviack's theory, our approach predicts a random pinning glass transition 
scenario that is qualitatively the same as the scenario obtained using a mean-field 
analysis of the spherical $p$-spin model and a mean-field version of the
random first-order transition theory.  
We use our approach to calculate quantities which are 
often considered to be indicators of growing dynamic correlations and static 
point-to-set correlations. We find that the so-called static overlap 
is dominated by the simple, low pinning fraction contribution. 
Thus, at least for randomly pinned fluid systems, 
only a careful quantitative analysis of simulation results can reveal
genuine, many-body point-to-set correlations.
\end{abstract}
\maketitle

Recently there have been several theoretical and simulational studies of
glassy dynamics of fluid systems in which some particles, randomly selected out 
of an equilibrium configuration, have been frozen or pinned 
\cite{KimEPL,KMSJPCM,CBPNAS,KrakPRE2011,BKPRE2012,KLPPhysica,CBEPL}. Originally these 
so-called partially pinned systems were considered to be just one special example 
of a broad class of model porous systems known as quenched-annealed 
binary mixtures \cite{MG,KrakPRE2010,KCKPRE2010}. 
However, it has now been realized that \emph{glassy} partially pinned systems 
can be used to reveal still unresolved aspects of the glass transition. 

First, it was proposed that by analyzing systems in which some particles,
taken out of an equilibrium configuration, have been frozen, one can study 
a growing ``amorphous order'' that is supposed to 
develop in glassy fluids \cite{BBJCP2004}. 
In early studies this idea was implemented using the so-called cavity geometry:
all particles except those within a spherical cavity were frozen and the 
local overlap of the original equilibrium configuration with configurations
equilibrated in the presence of pinned particles was monitored
\cite{BBCGVNatureP}. It was argued that the dependence of 
this overlap on the cavity diameter reveals a length characterizing the
so-called static point-to-set correlations, \textit{i.e.} 
correlations between the density at the center of the cavity (point) and 
the positions of the frozen particles (set). It was shown \cite{MSJSP} that, at least 
in simple models, these point-to-set correlations grow with increasing 
relaxation time. Subsequently, other geometries 
were introduced: one in which all particles except ones in a layer are frozen 
(the sandwich geometry), one in which all particles in a semi-infinite
space are frozen (the wall geometry) \cite{KRVBNatureP} or one in which  
a randomly chosen subset of particles, distributed uniformly throughout the 
system is frozen. It has been argued \cite{BKPRE2012} that the last geometry,
\textit{i.e.} the partially pinned system, is the best candidate to study 
growing static correlations. 

The second motivation for the recent interest 
in partially pinned systems comes from the realization that pinned particles, while 
maintaining the equilibrium structure of the fluid \cite{SKBJPC,KrakPRE2010},
may induce an ideal glass transition at temperatures or densities that are
more accessible to computer simulation studies \cite{CBPNAS}. Thus, the analysis 
of this so-called random pinning glass transition \cite{CBEPL} could both shed
light on the glass transition itself and provide a new way to test diverse
theoretical descriptions used to describe it. In particular, Krakoviack 
\cite{KrakPRE2011} recently argued that two different approaches, 
the mode-coupling theory \cite{Goetze} and the
random first order theory (RFOT) \cite{RFOT} 
make strikingly different predictions for the glassy behavior
of partially pinned systems. Thus, a simulational study of glassy behavior of a 
partially pinned system could easily disprove one of these approaches. 

The conclusion reached in Ref. \cite{KrakPRE2011} was surprising for two
reasons. First, it is usually assumed that mode-coupling and RFOT theories  
are related. In particular, the former theory is supposed to describe rather 
well the onset of glassy behavior while the RFOT approach is supposed to replace 
the mode-coupling transition with a crossover and to describe properties
of deeply supercooled fluids. It has to be admitted that recent theoretical 
analyzes of these theories in higher dimensions revealed some rather disturbing 
discrepancies \cite{Schilling,IMPRL,CIPZPRL}. In spite of this
fact, \emph{qualitative} disagreement between them in three dimensions
was not expected. Second, Cammarota and Biroli \cite{CBEPL} showed 
that a mean-field analysis
of the $p$-spin model with partially pinned spins predicts qualitatively the
same random pinning glass transition scenario as the mean-field version of the 
RFOT theory. Thus, we are now faced with a rather unpleasant qualitative disagreement 
between two mean-field-like calculations, mode-coupling theory and the mean-field
$p$-spin model. This disagreement is even more striking if 
we recall that (in the case of an un-pinned system) 
the so-called schematic model of mode-coupling theory is identical to 
the $p$-spin model (for $p=3$ used in Ref. \cite{CBEPL}).

Faced with the above described conundrum, we shall recall that there
is some freedom in the formulation of the mode-coupling approach, especially 
for more complex systems like mixtures and partially pinned systems.
Our goal in this Letter is to propose a simple, alternative mode-coupling approach 
that results in the random pinning glass transition scenario which is 
qualitatively consistent with the mean-field analysis of both 
the $p$-spin model \cite{CBEPL} and the RFOT theory \cite{CBPNAS}.
In addition, we will use our approach to calculate quantities that have been used
in earlier simulational studies to monitor the growth of dynamic correlations 
and static point-to-set correlations. We will show that a careful analysis of
these quantities is required to reveal genuine many-body effects.

A mode-coupling theory for a partially pinned system can be derived in
different ways. Here, we will outline a projection operator derivation 
which is easily compared with Krakoviack's theory \cite{KrakPRE2011,Krakder}.
Also, following Refs. \cite{KrakPRE2011,Krakder} we will refer to 
the mobile (un-pinned) particles as the fluid particles and to the pinned
ones as the matrix particles.
The fundamental dynamical variable used in our theory is the Fourier transform
of the microscopic fluid density,
%\begin{equation}\label{denfl}
$n^f(\mathbf{q}) = \sum_{j=1}^{N_f} e^{-i\mathbf{q}\cdot\mathbf{r}_j}$,
%\end{equation}
where $\mathbf{r}_j$ denotes the position of the $j$th fluid particle and $N_f$ is the
number of fluid particles. To describe the 
dynamics of the system, we use the fluid intermediate scattering function,
%\begin{equation}\label{Fdef}
$F(q;t) = 
\frac{1}{N_f}\left< n^f(\mathbf{q}) \exp(\Omega t) n^f(-\mathbf{q}) \right>$,
%\end{equation}
with $\Omega$ being the system's evolution operator and $\left< ... \right>$ 
denoting the average over all (fluid and matrix) particles of the system.
For simplicity, we assume here 
that the microscopic dynamics is Brownian, thus $\Omega$ is 
the many-body Smoluchowski operator. 
To derive an equation of motion for $F(q;t)$ we follow the standard 
procedure \cite{Goetze,SL} and arrive at an exact but formal memory function 
representation for $F(q;t)$ \cite{caveat}. We project the so-called fluctuating
force on the space spanned by the products of 
the fluid densities, $n^f(\mathbf{k})n^f(\mathbf{q}-\mathbf{k})$, and the products
of the fluid and matrix densities,  
$n^f(\mathbf{k})n^m(\mathbf{q}-\mathbf{k})$, where 
$n^m(\mathbf{q}) = \sum_{j=1}^{N_m} e^{-i\mathbf{q}\cdot\mathbf{s}_j}$ is 
the density of the matrix (pinned) particles, with $\mathbf{s}_j$ 
being the position of the $j$th matrix particle and $N_m$ being the number of the matrix
particles. Finally, we factorize the resulting four-point 
correlation functions \cite{der}.

We obtain the following equation of motion for $F(q;t)$, 
\begin{eqnarray}\label{MCT1} 
\lefteqn{ \int_0^t du \left(\delta(t-u)+ M^{\text{irr}}(q;t-u) \right) \partial_u F(q;u)
= }
\nonumber \\ &&
D q^2 n_f n_m c(q) h(q) - D q^2 (1 - n_f c(q) ) F(q;t).
\end{eqnarray} 
Here $D$ is the diffusion coefficient of an isolated fluid particle, 
$n_f$ and $n_m$ are the  densities of the fluid and matrix particles, respectively.
$h(q)$ and $c(q)$ are the
Fourier transforms of the complete system's 
correlation function and direct correlation function, respectively \cite{corrpinned}. 
Finally, $M^{\text{irr}}(q;t)$ is the irreducible memory function, 
\begin{eqnarray}\label{mem1}
M^{\text{irr}}(q;t) &=& \int \frac{d\mathbf{k}}{(2\pi)^3 }
\left[V^{(2)}(\mathbf{q},\mathbf{k}) F(k;t) F(|\mathbf{q}-\mathbf{k}|;t) 
\right. \nonumber \\ && \left.
+ V^{(1)}(\mathbf{q},\mathbf{k}) F(k;t) + V^{(0)}(\mathbf{q},\mathbf{k})
\right]
\end{eqnarray}
where
\begin{eqnarray}\label{V2}
V^{(2)}(\mathbf{q},\mathbf{k}) = 
\frac{n_f D}{2} \left[\hat{\mathbf{q}}\cdot\mathbf{k}c(k) 
+ \hat{\mathbf{q}}\cdot(\mathbf{q}-\mathbf{k})c(|\mathbf{q}-\mathbf{k}|)\right]^2
\end{eqnarray}
\begin{eqnarray}\label{V1}
&& V^{(1)}(\mathbf{q},\mathbf{k}) = n_m D 
\left[\hat{\mathbf{q}}\cdot(\mathbf{q}-\mathbf{k})c(|\mathbf{q}-\mathbf{k}|)\right]^2
\nonumber \\ && \times
\left[ 1+ n_m h(|\mathbf{q}-\mathbf{k}|) \right]
+ 2 n_f n_m D \hat{\mathbf{q}}\cdot(\mathbf{q}-\mathbf{k})c(|\mathbf{q}-\mathbf{k}|)
\nonumber \\ && \times 
\left[\hat{\mathbf{q}}\cdot\mathbf{k}c(k) 
+ \hat{\mathbf{q}}\cdot(\mathbf{q}-\mathbf{k})c(|\mathbf{q}-\mathbf{k}|)\right]
h(|\mathbf{q}-\mathbf{k}|)
\end{eqnarray}
\begin{eqnarray}\label{V0}
V^{(0)}(\mathbf{q},\mathbf{k}) &=& D n_m^2 \hat{\mathbf{q}}\cdot\mathbf{k}c(k)
\hat{\mathbf{q}}\cdot(\mathbf{q}-\mathbf{k})c(|\mathbf{q}-\mathbf{k}|)
\nonumber \\ && \times
(1-x) h(k)h(|\mathbf{q}-\mathbf{k}|)
\end{eqnarray}
In Eq. \eqref{V0}, $x$ is the fraction of particles that are pinned, $x=n_m/n$,
with $n=n_f+n_m$ being the total density of the system 
(in Refs. \cite{CBPNAS,BKPRE2012,CBEPL} the pinning fraction is denoted by $c$;
we changed the notation to avoid confusing the pinning fraction and
the direct correlation function).

It should be noted that there is a time-independent term at the right-hand-side
of Eq. \eqref{MCT1}
%the equation of motion \eqref{MCT1} 
and there is also a time-independent
contribution to $M^{\text{irr}}$ 
%the memory function 
originating from $V^{(0)}$, Eq. \eqref{V0}. 
These two terms make the long-time limit of $F(q;t)$ non-zero even below the 
glass transition.

Before turning to the results obtained from Eqs. 
(\ref{MCT1}-\ref{V0}) 
we compare our present approach to 
Krakoviack's \cite{KrakPRE2011,Krakder}. Krakoviack also follows the projection
operator procedure detailed in Ref. \cite{Goetze}. 
However, he takes as the fundamental dynamical variable the so-called
relaxing part of the fluid density,
$\delta n^f(\mathbf{q}) =   n^f(\mathbf{q}) - \left<n(\mathbf{q})\right>^f$ 
where $\left< ... \right>^f$ denotes the 
equilibrium average over the positions of the fluid particles only, with 
the matrix (pinned) particles treated as a set of fixed obstacles. Furthermore,
he projects the fluctuating force on the space spanned by the products of 
the relaxing fluid densities, 
$\delta n^f(\mathbf{k})\delta n^f(\mathbf{q}-\mathbf{k})$, the products
of the relaxing fluid density and the matrix density, 
$\delta n^f(\mathbf{k})n^m(\mathbf{q}-\mathbf{k})$, 
and the products of the relaxing fluid density and the average fluid density for 
a given set of positions of the matrix particles, 
$\delta n^f(\mathbf{k})\left<n(\mathbf{q}-\mathbf{k})\right>^f$. 
It should be noted that, while the all the variables that we use are 
either one-particle additive or pairwise-additive, the average fluid density
$\left<n(\mathbf{q})\right>^f$ used in Ref. \cite{Krakder} 
is \emph{not} one-particle or pairwise additive
in terms of the matrix particles; instead, it includes terms involving arbitrarily
many matrix particles. 

The main difference between our approach and that of Refs. \cite{KrakPRE2011,Krakder} 
is that we use the mode-coupling approach to find both the time evolution of the
relaxing density fluctuations and the average non-relaxing density fluctuations. 
In our language the latter quantity is the long-time limit of the fluid intermediate
scattering function, $F(q)\equiv\lim_{t\to\infty} F(q;t)$. In contrast, Krakoviack 
uses the mode-coupling approach only to find the time evolution of the
relaxing density fluctuations and resorts to  
a separate static approach, replica OZ integral equation theory \cite{MG,KrakPRE2010},
to calculate the properties of the non-relaxing density fluctuations. In his 
language the latter fluctuations are characterized by the so-called disconnected or 
blocked fluid structure factor, 
$S^b(q)=\frac{1}{N_f} 
\overline{\left<n(\mathbf{q})\right>^f \left<n(-\mathbf{q})\right>^f}$
where $\overline{ \cdots }$ denotes the average over the
positions of the pinned particles. 

Two arguments were put forward for the approach used in Ref. \cite{Krakder}. First,
it was argued that since the calculation of the properties of non-relaxing fluid 
density fluctuations is a static problem, it is appropriate to use the mode-coupling 
approach for the dynamics of the relaxing part only. Second, in the formulation
of Ref. \cite{Krakder} the so-called connected fluid structure factor,
$S^c(q)=\frac{1}{N_f} \left< \delta n^f(\mathbf{q})\delta n^f(-\mathbf{q}) \right>$,
appears naturally in the expression for the so-called characteristic frequency 
\cite{Krakder} which, for a Brownian system, is given by $\omega^2(q) = Dq^2/S^c(q)$. 

To answer these arguments we note that the mode-coupling calculation of the non-decaying 
part of the fluid intermediate scattering function $F(q;t)$ can be given 
a static interpretation \cite{GSEPL}. 
Thus, our approach can be considered to be a combination
of a specific static calculation that is consistent with the mode-coupling prediction for
the non-decaying part of $F(q;t)$ and the
mode-coupling calculation of the decaying part of $F(q;t)$. Moreover, if the
equation of motion is re-written in terms of the decaying part only,
the characteristic frequency acquires the form $\omega^2(q) = Dq^2/(S^f(q)-F(q))$,
where $S^f(q) = F(q;t=0) = 1+n_f h(q)$ is the fluid static structure factor. This form,
upon identification of $F(q;\infty)$ and $S^b(q)$, is identical to that obtained
in Ref. \cite{Krakder}.

Eqs. (\ref{MCT1}-\ref{V0}) allow us to calculate 
both $F(q)$, \textit{i.e.} the non-decaying long time limit of $F(q;t)$, and the
time dependence of $F(q;t)$. 
In fact, for the fluid states the small $x$ limit of $F(q)$ can be easily obtained 
from Eq. \eqref{MCT1},
\begin{equation}\label{smallx}
F(q) = x  n^2 h^2(q) + O(x^2).
\end{equation}
It can be showed that this result coincides with the exact small $x$ limit
of the blocking structure factor, $S^b(q)$. 

To solve Eqs. (\ref{MCT1}-\ref{V0}) numerically we disctretized the wave-vector space
using 500 wave-vectors with the smallest one equal to $0.05$ and a 
spacing of $0.1$. 
The resulting set of coupled integro-differential equations was solved using
the procedure outlined in Ref. \cite{FSmctsol}. We used the hard sphere interaction
potential and the Percus-Yevick approximation for the static correlations,
$h(q)$ and $c(q)$. The results below are presented using reduced units: 
distance is measured in terms of the hard sphere diameter $\sigma$ and 
time in terms of $\sigma^2/D$. 

\begin{figure}
\includegraphics[width=3.4in]{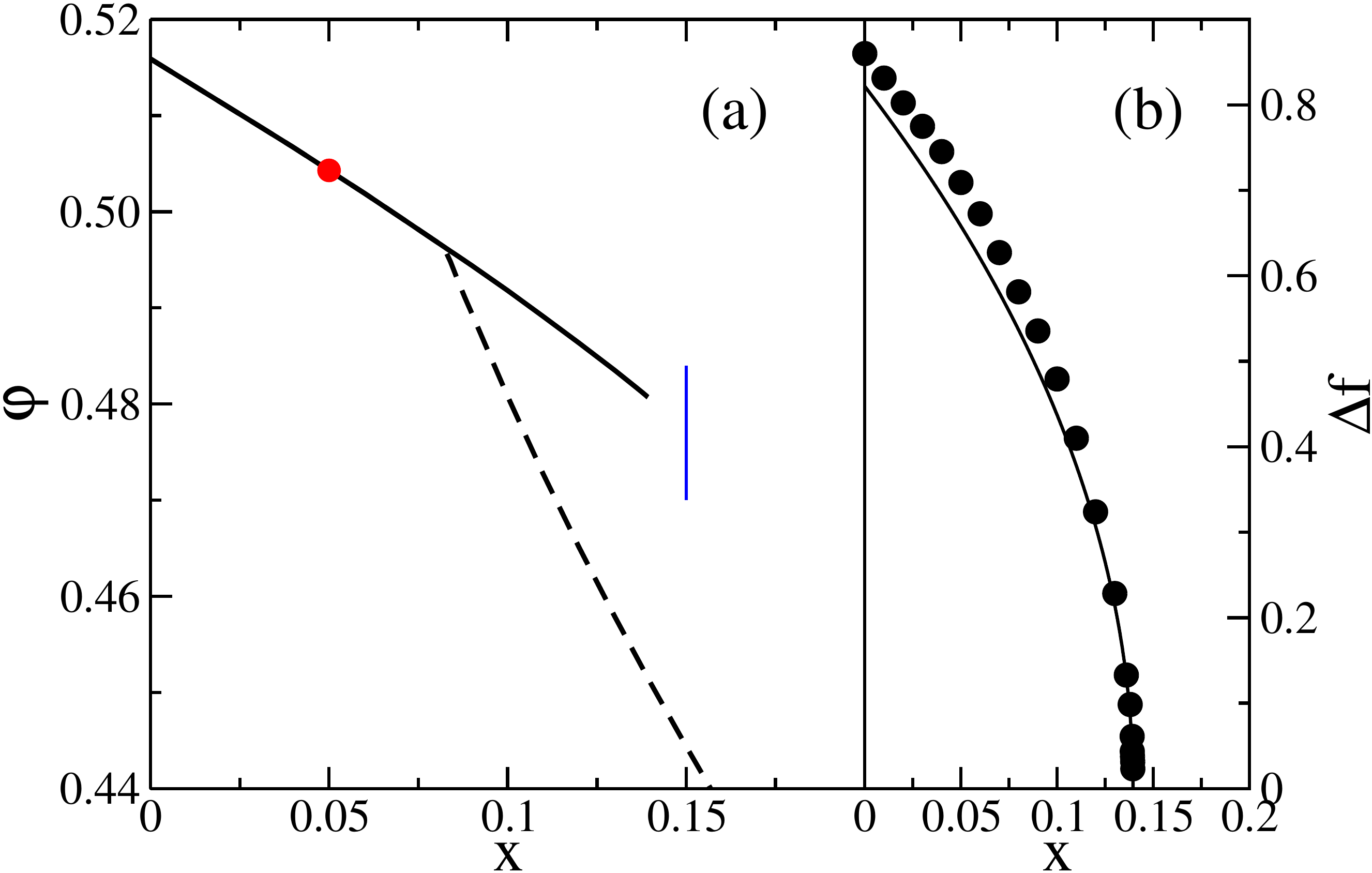}\\
\caption{\label{fig:pd} (a) dynamic phase diagram of 
a partially pinned hard sphere system. $x$ and $\varphi$ denote, respectively, 
the fraction of particles that are pinned and the volume fraction of all 
(mobile and pinned) particles. Solid line: discontinuous mode-coupling transition.
Dashed line: localization transition (\textit{i.e.}  
vanishing self-diffusion coefficient) \cite{sd}. The filled circle
and the thin vertical line indicate states for which the time dependence of the 
intermediate scattering function is shown in Fig. \ref{fig:timedep}. 
(b) Circles: $x$ dependence of the discontinuity of the long-time limit of the 
normalized intermediate scattering function, $f(q)=F(q)/F(q;t=0)$,  along the 
mode-coupling transition line shown in (a); 
thin solid line shows the square root singularity, 
$\Delta f \propto (x-x_c)^{1/2}$.}
\end{figure}

\begin{figure}
\includegraphics[width=3.4in]{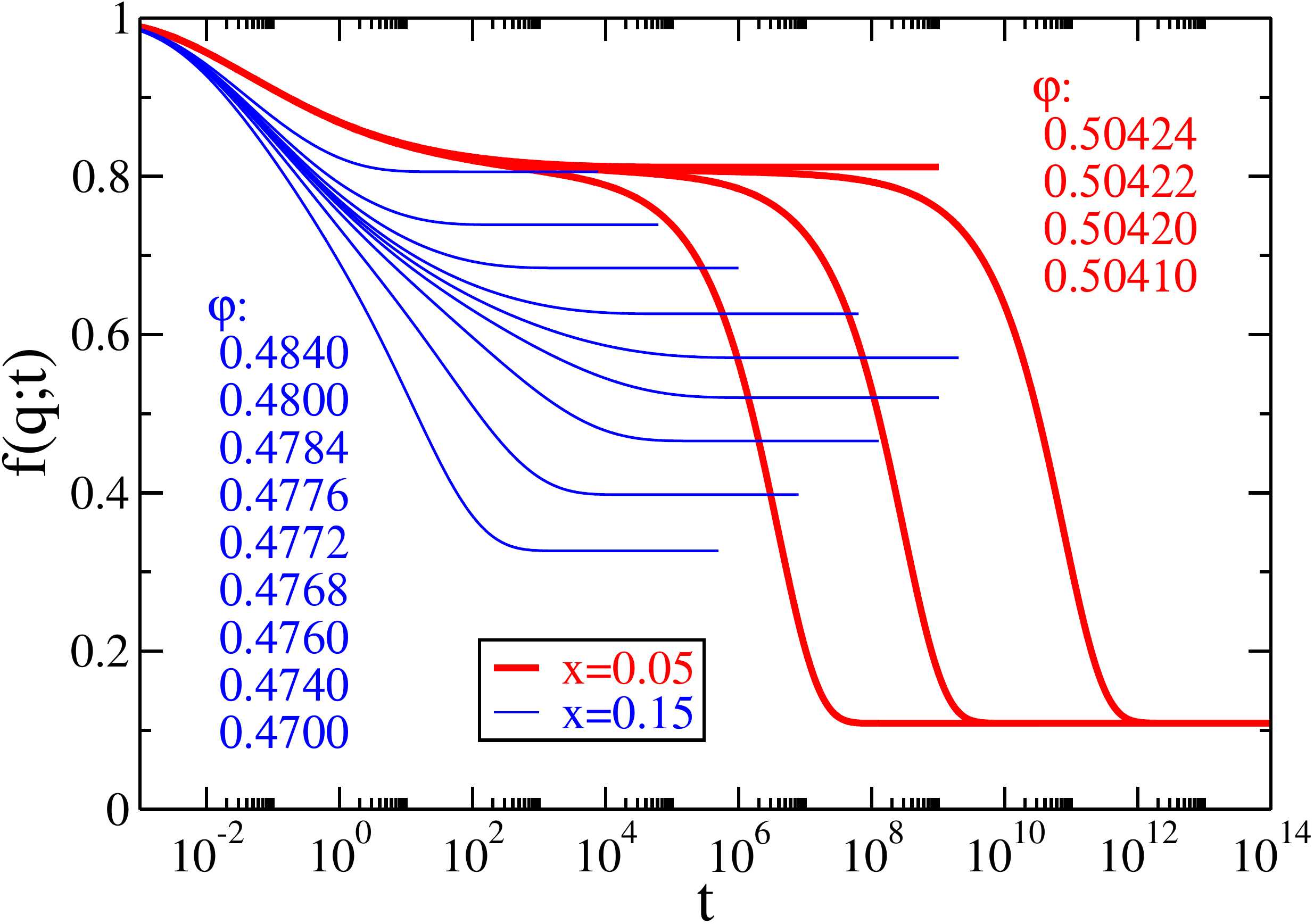}\\
\caption{\label{fig:timedep} Time dependence of the normalized intermediate scattering 
function, $f(q;t)=F(q;t)/F(q;t=0)$. Thick solid lines: $x=0.05$ and 
$\varphi$ near the mode-coupling
transition, solid circle in the
left panel of Fig. \ref{fig:pd}. Thin solid lines: 
$x=0.15$ and $\varphi$ along the thin vertical line in Fig. \ref{fig:pd}a.} 
\end{figure}

The random pinning glass transition scenario is presented in 
Figs. \ref{fig:pd} and \ref{fig:timedep}. 
The standard mode-coupling transition present at $x=0$ extends into the $x>0$ 
plane. However, the discontinuity of $F(q)$ at this transition decreases
with increasing $x$ and the transition disappears at $x_c\approx 0.1395$. 
Beyond $x_c$, the intermediate scattering function changes continuously with 
the volume fraction $\varphi = n\pi/6$. This scenario is qualitatively
consistent with the $p$-spin model results presented in Ref. \cite{CBEPL} 
and with the mean-filed RFOT analysis presented in \cite{CBPNAS}. Thus,
it agrees with the striking prediction of Cammarota and Biroli that it is possible
to enter the glass phase without ever encountering a divergence of the relaxation time. 

We used the numerical solution of Eqs. (\ref{MCT1}-\ref{V0})
to calculate the so-called collective and single-particle overlap functions. 
To define these functions we followed Berthier and Kob 
\cite{BKPRE2012}. Specifically, $Q(t)$ is proportional to the 
the probability that a spherical
cell of radius $a$ is occupied at both $t=0$ and $t\to\infty$, 
\begin{eqnarray}\label{Qcoll}
Q(t) \!\!\! &=& \!\!\! \frac{3}{4\pi a^3} \int d\mathbf{r}_1 \int d\mathbf{r}_2
F(|\mathbf{r}_1-\mathbf{r}_2|;t) \theta(a-r_1) \theta(a-r_2)
\nonumber \\ && + Q_{\text{rand}}
\end{eqnarray}
where $F(r;t)$ is the inverse Fourier transform of $F(q;t)$ and
$Q_{\text{rand}}=4\pi n a^3/3$. In the definition of 
the single-particle overlap $Q_{\text{self}}(t)$, 
$F(r;t)$ is replaced by the inverse Fourier transform of of the self-intermediate
scattering function $F^s(q;t)$ \cite{sd} and $Q_{\text{rand}}$ is absent. 
To make connection with Ref. \cite{BKPRE2012} we chose $a=0.3$. 

\begin{figure}
\includegraphics[width=3.4in]{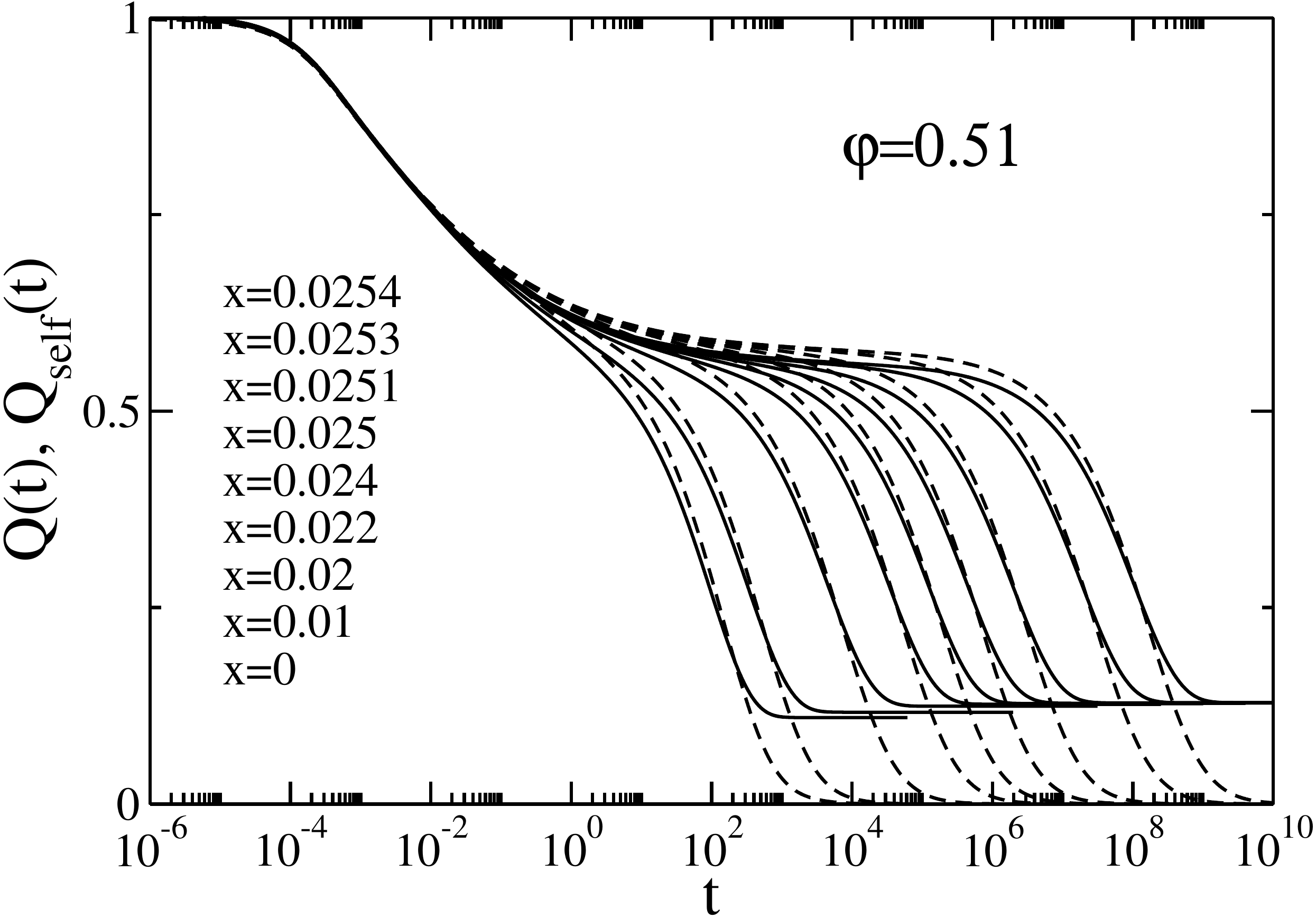}\\
\caption{\label{fig:Qt} Time dependence of the collective overlap
$Q(t)$ (solid lines) 
and the single particle overlap $Q_{\text{self}}(t)$ (dashed lines) for $\varphi=0.51$
and values of $x$ indicated in the figure.} 
\end{figure}

In Fig. \ref{fig:Qt} we show the time-dependence of
the collective and single-particle overlaps for $\varphi=0.51$.
For smaller $x$ our predictions are qualitatively similar to computer simulation results 
showed in
Fig. 2(c) of Ref. \cite{BKPRE2012}. However, upon approaching the mode-coupling
transition $Q(t)$ predicted by the theory shows a classic 
mode-coupling-like two step decay whereas the simulation results 
exhibit a continuous increase of both 
the intermediate-time plateau and the long-time plateau. 

\begin{figure}
\includegraphics[width=3.4in]{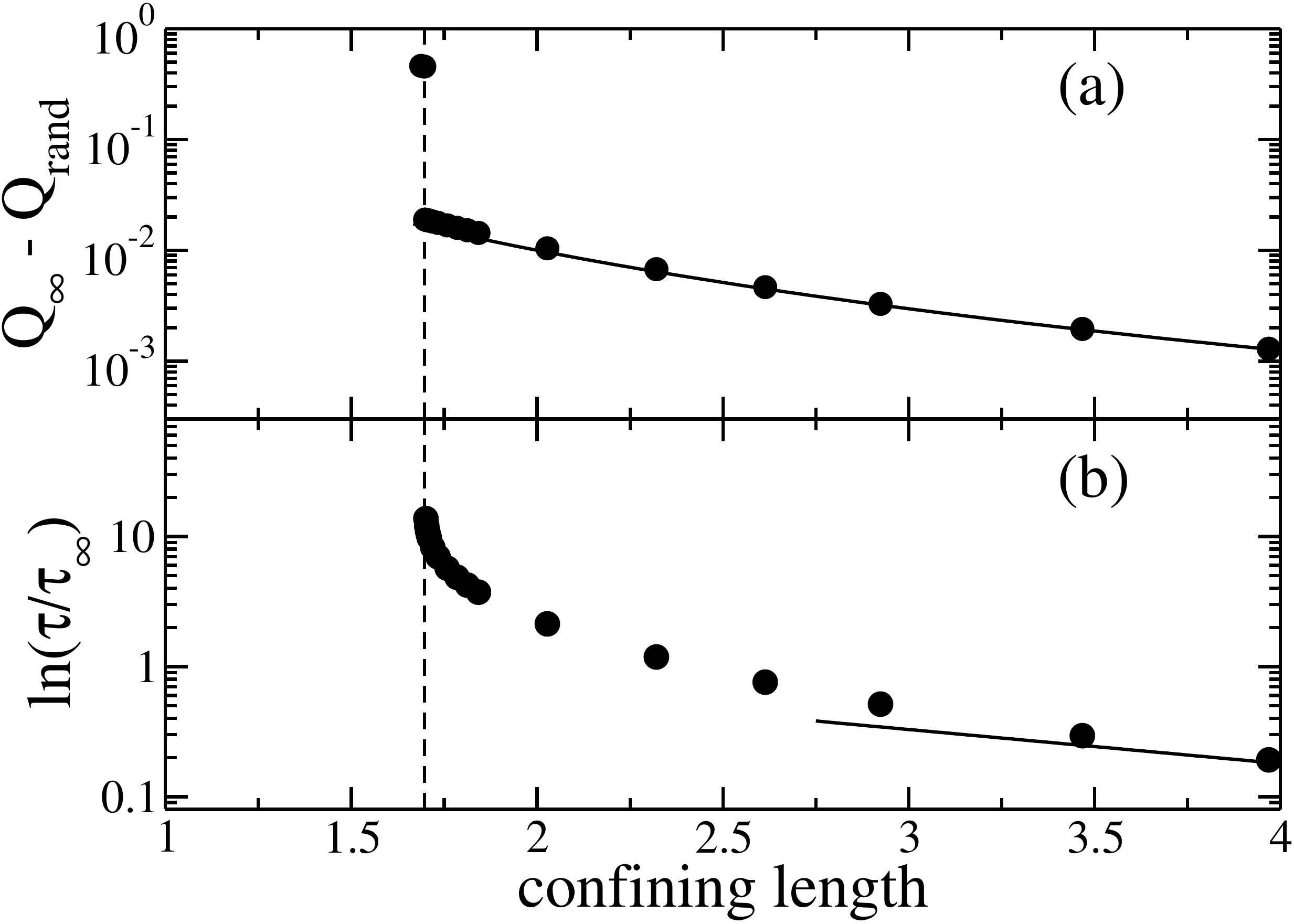}\\
\caption{\label{fig:length} (a) Dependence of the overlap $Q_\infty$ 
on the confining length $l=0.5 x^{-1/3}$ for $\varphi = 0.51$.  
Solid line indicates the overlap 
originating from the small $x$ limit of $F(q)$, Eq. \eqref{smallx}.
(b) Dependence of the single-particle overlap relaxation time 
on the confining length $l=0.5 x^{-1/3}$. Solid line shows the exponential
dependence $\ln(\tau/\tau_{\infty})\propto \exp(-l/\xi_{\text{IMCT}})$
with $\xi_{\text{IMCT}}=1.66$ for $\varphi=0.51$ \cite{SFPRE2010}.
Dashed vertical line through (a) and (b) indicates the
mode-coupling transition.} 
\end{figure}

In Fig. \ref{fig:length}a we show the length dependence of the so-called
static overlap, \textit{i.e.} the nontrivial part of the
long time plateau of the collective overlap, $Q_{\infty}-Q_{\text{rand}}$, where
$Q_{\infty}=\lim_{t\to\infty} Q(t)$. The length, $l=0.5 x^{-1/3}$,  
is the so-called confining length \cite{BKPRE2012}. 
Perhaps fortuitously, for the range of lengths corresponding to
non-glassy states, the values of $Q_\infty-Q_{\text{rand}}$ are quite close
to those obtained from computer simulations. However, according to 
our mode-coupling approach an increase of $Q_\infty-Q_{\text{rand}}$ is 
followed by a discontinuity
at the mode-coupling transition line whereas the values obtained from simulations
increase continuously with decreasing $l$. Importantly, growing values of 
$Q_\infty-Q_{\text{rand}}$ are often associated with growing amorphous order and
more specifically growing static point-to-set correlations. As indicated in
Fig. \ref{fig:length}a (and also noted in Ref. \cite{BKPRE2012}), 
for the range of lengths corresponding to
non-glassy states,  $Q_\infty-Q_{\text{rand}}$ is
dominated by the small $x$ contribution that originates from expression \eqref{smallx}.
This contribution has a rather trivial origin and it should not be associated
with any many-body correlations \cite{CCT}. 
%Thus, an analysis of simulational results for the 
%partially pinned systems should subtract the baseline originating from
%Eq. \eqref{smallx}. 
It would be interesting to investigate whether there is
a similar simple ``baseline'' contribution for other geometries considered in Ref. 
\cite{BKPRE2012}. Finally, in Fig. \ref{fig:length}b we show the length dependence 
of the relaxation time $\tau$ defined through the single particle overlap, 
$Q_{\text{self}}(\tau) = e^{-1}$. We find that our mode-coupling approach
overestimates the influence of the pinning on the relaxation time: the values
of $\ln(\tau/\tau_{\infty})$, where $\tau_\infty$ is the relaxation time of
the $l=\infty$ (\textit{i.e.} un-pinned, $x=0$)
system, are consistently above those obtained from computer 
simulations. We note that at large $l$, the length of the relaxation time
seems to become consistent with the value of the dynamic correlation length obtained 
from the so-called inhomogeneous mode-coupling theory (IMCT)
\cite{BBMRPRL2006,SFPRE2010}. 

In summary, we proposed an alternative mode-coupling approach for partially pinned
systems. Our approach agrees qualitatively with the mean-field analysis of
the $p$ spin model and with the results obtained from the RFOT theory. 
We showed that the contribution to the long-time limit of the collective overlap 
originating from the presence 
of pinning is dominated by the trivial small $x$ limit. 
This small $x$ contribution is not related to any growing point-to-set correlations.

We gratefully acknowledge the support of NSF Grant CHE 0909676. 
GS thanks G. Biroli and 
C. Cammarota for a discussion that inspired this work. We thank 
G. Biroli, L. Berthier, C. Cammarota and V. Krakoviack for comments on
the manuscript.


\begin{thebibliography}{99}
\bibitem{caveat} While following the procedure of Refs. \cite{Goetze,SL} 
it is important to 
remember that the fluid density has non-vanishing correlations with the matrix density.
\bibitem{der} The details of the derivation will be presented elsewhere.
\bibitem{CCT} A similar conclusion was reached in Ref. \cite{CCTPRL}.
\end{thebibliography}
\end{document}